\documentclass[a4paper,11pt]{article}
\usepackage{graphicx,amssymb,bm,latexsym,color,epsf}
\makeatletter
\@addtoreset{equation}{section}

\makeatother
\newcommand{\bea}{\begin{eqnarray}}
\newcommand{\ena}{\end{eqnarray}}
\newcommand{\vs}[1]{\vspace{#1 mm}}
\newcommand{\hs}[1]{\hspace{#1 mm}}
\renewcommand{\a}{\alpha}
\renewcommand{\b}{\beta}
\renewcommand{\c}{\gamma}
\newcommand{\G}{\Gamma}
\renewcommand{\d}{\delta}

\newcommand{\s}{\sigma}
\renewcommand{\t}{\theta}

\newcommand{\la}{\lambda}

\newcommand{\nn}{\nonumber\\}
\newcommand{\p}[1]{(\ref{#1})}

\newcommand{\lra}{\leftrightarrow}
\newcommand{\br}{\bar R}
\newcommand{\bg}{\bar g}

\newcommand{\tg}{\tilde g}
\newcommand{\bk}{{\bf k}}

\begin{document}

\begin{titlepage}

\begin{flushright}
KU-TP 063 \\
%\today
\end{flushright}

\vs{10}
\begin{center}
{\Large\bf Ultraviolet Fixed Points in Conformal Gravity \\
and General Quadratic Theories}
\vs{15}

{\large
Nobuyoshi Ohta\footnote{e-mail address: ohtan@phys.kindai.ac.jp}$^{,a}$
and Roberto Percacci\footnote{e-mail address: percacci@sissa.it}$^{,b,c}$
} \\
\vs{10}

$^a${\em Department of Physics, Kinki University,
Higashi-Osaka, Osaka 577-8502, Japan}

$^b${\em International School for Advanced Studies, via Bonomea 265, 34136 Trieste, Italy}

$^c${\em INFN, Sezione di Trieste, Italy}

\vs{15}
{\bf Abstract}
\end{center}

We study the beta functions for four-dimensional conformal gravity using two different
parametrizations of metric fluctuation, linear split and exponential parametrization.
We find that after imposing the traceless conditions, the beta functions are the same
in four dimensions though the dependence on the dimensions are quite different.
This indicates the universality of these results.
We also examine the beta functions in general quadratic theory with the Einstein and
cosmological terms for exponential parametrization, and find that it leads to
results for beta functions of dimensionful couplings different from linear split,
though the fact that there exists nontrivial fixed point remains the same
and the fixed points also remain the same.

\end{titlepage}
\newpage
\setcounter{page}{2}

\section{Introduction}

It has long been known that Einstein gravity theory with quadratic curvature terms are
renormalizable~\cite{Stelle}.
Early attempts at computing beta functions are made in~\cite{JT,ft1}, and
the correct beta functions for the dimensionless couplings have been obtained by
Avramidi and Barvinsky~\cite{Av}.
The theories are shown to be asymptotically free~\cite{ft1}.
However it is known that this class of theories
suffer from the problem of ghosts and cannot give healthy physical theories.
For this reason, the theories are not taken seriously as a candidate of quantum gravity.

Recently, however, it has been shown that a special class of such theories can be unitary
in three dimensions. These are called new massive gravity~\cite{BHT1}.
The classification of all possible unitary theories in three dimensions with quadratic
curvature terms is given in \cite{Ohta1}.
With quadratic curvature terms, it is expected that these theories may give
unitary and renormalizable quantum gravity. If the unitary theories are also renormalizable,
this would be the first example of complete theory of quantum gravity although in three dimensions.
Unfortunately it turns out that the unitarity and renormalizability are just incompatible
with each other in this class of three-dimensional theories~\cite{deser,BHT2,Ohta2}.
Given this result, it seems that the only possible way to make sense of these theories
is to resort to the notion of asymptotic safety~\cite{Weinberg}--\cite{CPR}.
These are those theories which have nontrivial ultraviolet fixed points and could provide
consistent UV completion of the theories within Wilson's formulation of renormalization.
Asymptotic safety may be as predictive as asymptotic freedom.

In our previous papers~\cite{Ohta3,OP} we have studied this class of theories in diverse dimensions
and derived beta functions for the coupling constants in the theories.
We have found that there are certainly nontrivial fixed points as well as Gaussian fixed points
in all dimensions studied, thus confirming and extending earlier results in
four dimensions~\cite{Av,lauscher}--\cite{AKKLR}.
We have also studied the theories near four dimensions, and found that there is an additional
fixed point slightly away from four dimensions, but this fixed point disappears just on
the dimension four. We suggested that this additional fixed point may corresponds to
a Weyl-invariant theory because at that point the coefficient of the Weyl-non-invariant term
vanishes, and the reason that this fixed point disappear in $D=4$ is attributed to the fact
that our approach did not take into account the Weyl invariance of the theory at that fixed
point; we did not impose that the trace of the metric fluctuation vanish.
This is all right outside $D=4$ since there is no such invariance, but becomes a problem in $D=4$.
However the fixed point value was slightly different from the one computed for conformal
gravity in earlier paper~\cite{BS1}. It is one of the purposes of this paper to further study
the fixed points in the conformal theories.

Another related issue is the gauge- and parametrization-dependence of the fixed points.
In all attempts at the quantum treatment of gauge theories, we have to fix the gauge, and
it is an important issue to know if and how the physical result depends on the gauge.
In the gravity theory, there is also a problem how to parametrize the fluctuation
around backgrounds. Here we also study how our result may change
if we use different parametrization of the metric.
Namely we calculate the beta functions in $D=4$ theory with Weyl curvature squared together
with Gauss-Bonnet terms, by imposing the traceless condition. The calculation is
done in the standard linear split of the metric fluctuation:
\bea
g_{\mu\nu}= \bg_{\mu\nu} + h_{\mu\nu}\ .
\label{linear}
\ena
We then find that the result agrees with the earlier one.
We also evaluate beta functions in a new parametrization of nonlinear exponential
type~\cite{Kawai}:
\bea
g_{\mu\nu}= \bg_{\mu\rho} ( e^h )^\rho{}_\nu\ ,
\label{nonlinear}
\ena
in order to check if this is a universal result.

There are some advantage of using the latter parametrization:

(1) The metric in ~\p{nonlinear} is always positive for large fluctuation, whereas
this is not true for the linear split~\p{linear}.

(2) It helps to separate the gauge dependence of the obtained results~\cite{Benedetti}--\cite{PV}.

(3) It turns out that the unphysical singularities which often appear in the flow
equations can be avoided for the nonlinear parametrization~\p{nonlinear}~\cite{PV}.

It is thus interesting to check the result also in the new parametrization. We find that
the result does not change for dimensionless couplings, thus confirming the universality
of the beta functions in this case.
However, we find that the beta functions of the dimensionful couplings, the Newton constant
and cosmological terms, do change.
We discuss physical implications of these results.

This paper is organized as follows.
In Sec.~2, we present the theory we consider, and summarize our conventions.
In Sec.~3, we derive the quadratic part of our action for the linear split~\p{linear}
and make gauge fixing.
In Sec.~4, we show how the result in Sec.~3 is modified for the nonlinear
parametrization~\p{nonlinear}.
We also discuss the case of Einstein theory with cosmological constant and quadratic
terms in the curvature in the nonlinear parametrization~\p{nonlinear}.
This serves to check how the result in our previous paper~\cite{OP} changes for
different parametrization.
In Sec.~6, we briefly summarize the functional renormalization group equations (FRGE), and
then derive the beta functions in the above cases in Sec.~7.
Sec.~8 is devoted to our conclusions.

\section{The Action}

We consider a theory invariant under the Weyl transformation in four dimensions.
The action is given by
\bea
S = \int d^4 x \sqrt{- g} \Big[ \frac{1}{2\la} C^2 - \frac{1}{\rho} E \Big] ,
\label{action}
\ena
containing two dimensionless couplings $\la$ and $\rho$.
Here
\bea
C^2 = R_{\mu\nu\a\b}^2- 2 R_{\mu\nu}^2+ \frac{1}{3} R^2,
\ena
is the square of the Weyl tensor and
\bea
E = R_{\mu\nu\a\b}^2- 4 R_{\mu\nu}^2+ R^2,
\ena
is the Gauss--Bonnet combination, which is topological in four dimensions.
Nevertheless, we should keep this term in discussing renormalization of the theory.

The action~\p{action} is invariant under the Weyl rescaling
\bea
g_{\mu\nu}=\Omega^2 \tg_{\mu\nu},
\ena
with $\Omega$ a function of spacetime.

The action~\p{action} has the alternative form
\bea
S &=& \int d^4 x \sqrt{- g} \Big[ \a R^2 +\b R_{\mu\nu}^2 + \c R_{\mu\nu\rho\la}^2 \Big] ,
\label{action1}
\ena
where
\bea
\a = -\frac{1}{\rho}+\frac{1}{6\la}, ~~
\b = \frac{4}{\rho} -\frac{1}{\la}, ~~
\c = -\frac{1}{\rho} +\frac{1}{2\la}.
\label{coe}
\ena

We use the background field method and split metric into general backgrounds and quantum parts,
as given in \p{linear} and \p{nonlinear}.
The cutoff is constructed in such a way that the flow equation is invariant
under background transformations, both diffeos. and Weyl transformations.
Then we have to fix the gauge; we choose the gauge for the quantum Weyl transformations
by imposing the trace of the metric fluctuation is zero.
The final result should not depend on the gauge due to the Weyl invariance.

\section{Quadratic expansion of the action in the linear split}

In order to derive the effective action at the one-loop level,
or to calculate the one-loop beta functions,
we need the expansion of the action to second order in $h_{\mu\nu}$.
This calculation is discussed in detail in \cite{OP} for linear split~\p{linear}.
In this and next three sections, the result is presented for any dimensions $D$,
though our main concern in the present paper is $D=4$.

We first summarize the final form for the linear split~\p{linear}, where we have dropped
terms with linear derivatives acting on the fluctuation and terms with two derivatives
acting on a background curvature (omitting indices, these are of the form $h(\nabla\br)\nabla h$
and $h(\nabla\nabla\br)h$;
such terms do not contribute to the final results, see for example \cite{Av,BS2}).
The terms proportional to $\alpha, \b$ and $\c$ can be written in the form~\cite{OP,BC}
\bea
\a h^{\mu\nu}\hs{-7}&&  \Big[ \nabla_\mu \nabla_\nu \nabla_\a \nabla_\b
- \bg_{\mu\nu} \Box \nabla_\a \nabla_\b - \bg_{\a\b} \nabla_\mu \nabla_\nu \Box
+\bg_{\mu\nu} \bg_{\a\b} \Box^2 -\br \bg_{\nu\b} \nabla_\a \nabla_\mu \nn
&& - (2 \br_{\mu\nu} - \br \bg_{\mu\nu})\nabla_\a \nabla_\b
+ 2 \br_{\mu\nu} \bg_{\a\b}\Box +\frac12 \br (\bg_{\mu\a}\bg_{\nu\b}
- \bg_{\mu\nu}\bg_{\a\b})\Box -\br \br_{\a\b}\bg_{\mu\nu} \nn
&&
+ 2\br \br_{\mu\a} \bg_{\b\nu} + \br_{\mu\nu} \br_{\a\b}
-\frac{1}{4}J_{\mu\nu\alpha\beta}\br^2
\Big] h^{\a\b}, \nn
\b h^{\mu\nu}\hs{-7}&& \Big[ \frac12 \nabla_\mu \nabla_\nu \nabla_\a \nabla_\b
- \frac12 \bg_{\mu\nu} \nabla_\a \Box \nabla_\b
-\frac12 \bg_{\nu\b} \nabla_\mu \Box \nabla_\a +\frac14( \bg_{\mu\a} \bg_{\nu\b}
+\bg_{\mu\nu} \bg_{\a\b})\Box^2 +\frac12 \br_{\nu\b} \nabla_\a \nabla_\mu \nn
&& -2 \br_\mu^\rho \bg_{\nu\b} \nabla_\rho\nabla_\a
+ \frac32 \bg_{\a\b}\br_{\rho\mu} \nabla^\rho \nabla_\nu
+ \br_{\mu\a\nu\b} \Box +\frac14 (2 \bg_{\mu\a} \bg_{\nu\b}
- \bg_{\mu\nu} \bg_{\a\b}) \br^{\rho\la}\nabla_\rho \nabla_\la
-\frac32 \br^\rho_\b \br_{\rho\mu\nu\a} \nn
&& + \br_{\mu\rho\la\nu} \br_\a{}^{\rho\la}{}_\b
 +\frac12 \bg_{\nu\b} \br_{\mu\rho}\br^{\rho}_\a
- \bg_{\a\b} \br_{\mu\rho} \br^\rho_\nu
-\frac{1}{4}J_{\mu\nu\alpha\beta}\br_{\rho\sigma}^2
\Big] h^{\a\b}, \nn
\c h^{\mu\nu} \hs{-5}&& \hs{-2} \Big[ \nabla_\mu \nabla_\nu \nabla_\a \nabla_\b
+ \bg_{\mu\a} \bg_{\nu\b} \Box^2
-2 \bg_{\nu\b}\nabla_\mu \Box \nabla_\a
- 2 \bg_{\nu\b} \br_{\mu\rho\la\a} \nabla^\rho \nabla^\la
+ 3 \br_{\mu\a\nu\b}\Box \nn
&&
- 4 \br_{\mu \a\rho\nu} \nabla^\rho \nabla_\b
- 4 \bg_{\nu\b} \br_{\rho\mu} \nabla^\rho \nabla_\a
-2\bg_{\nu\b}\br_{\mu\a}\Box
+ \bg_{\mu\a} \bg_{\nu\b} \br_{\rho\la} \nabla^\rho \nabla^\la \nn
&& -2\bg_{\mu\nu} \br_{\rho\a\la\b}\nabla^\la\nabla^\rho
+ 4 \br_{\mu\a} \nabla_\nu \nabla_\b
+ 2 \bg_{\nu\b} \br_{\mu\la\rho\s} \br_\a{}^{\la\rho\s}
-2 \bg_{\nu\b} \br^{\rho\la} \br_{\mu\rho\a\la}
\nn
&&
- \br_{\mu\la\rho\b} \br_\nu{}^{\rho\la}{}_\a
+ 3 \br_{\mu\a}{}^{\rho\la} \br_{\nu\rho\b\la}
-3 \br_{\mu\la\nu\rho} \br_\a{}^{\rho\la}{}_\b
-3 \br^\rho_\a \br_{\mu\b\nu\rho}
\nn
&&
+2 \br_{\mu\a} \br_{\nu\b}
- \bg_{\mu\nu} \br_\a{}^{\rho\la\s} \br_{\b\rho\la\s}
-\frac{1}{4}J_{\mu\nu\alpha\beta}\br_{\rho\sigma\lambda\tau}^2
\Big] h^{\a\b}.
\label{r3}
\ena
Here and in what follows, a bar indicates that the quantity is evaluated on the general background;
the indices are raised, lowered and contracted by the background metric $\bg$,
the covariant derivative $\nabla$ is constructed with the background metric.
The tensor $J$ is defined by
\bea
J_{\mu\nu\a\b}=\d_{\mu\nu,\a\b}-\frac12 \bg_{\mu\nu} \bg_{\a\b},
\label{defJ}
\ena
where
\bea
\d_{\mu\nu,\a\b} = \frac12 (\bg_{\mu\a} \bg_{\nu\b} + \bg_{\mu\b} \bg_{\nu\a})
\equiv \hat 1,
\label{id}
\ena
is the identity in the space of symmetric tensors.
We should note that due to the presence of the external factors of $h$,
the expression in the square bracket is automatically symmetrized under the interchanges
$\mu\leftrightarrow\nu$, $\alpha\leftrightarrow\beta$ and
$(\mu,\nu)\leftrightarrow (\alpha,\beta)$.

The BRST transformation for the fields is found to be
\bea
\d_B g_{\mu\nu} \hs{-2}&=&\hs{-2} -\d \la [ g_{\rho\nu}\nabla_\mu c^\rho
 + g_{\rho\mu}\nabla_\nu c^\rho
] \equiv -\d\la{\cal D}_{\mu\nu,\rho}c^\rho, \nn
\d_B c^\mu \hs{-2}&=&\hs{-2} -\d\la c^\rho \nabla_\rho c^\mu,~~~
\d_B \bar c_\mu = i \d\la\, B_\mu, ~~~
\d_B B_\mu = 0,
\label{brst}
\ena
which is nilpotent. Here $c^\mu, \bar c_\mu$ and $B_\mu$ are the Faddeev-Popov ghost,
anti-ghost and an auxiliary field, respectively, and $\d\la$ is an anticommuting parameter.
The gauge fixing term and the Faddeev-Popov ghost terms are concisely written as
\bea
{\cal L}_{GF+FP}/\sqrt{-\bg} \hs{-2}&=&\hs{-2} i \d_B [\bar c_\mu Y^{\mu\nu}
(\chi_\nu-\frac{a}{2} B_\nu)]/\d\la
\nn \hs{-2}&=&\hs{-2}
 - B_\mu Y^{\mu\nu} \chi_\nu
+ i \bar c_\mu Y^{\mu\nu} ( \nabla^\la{\cal D}_{\la\nu,\rho}
+ b \nabla_\nu{\cal D}_{\la,\rho}^\la) c^\rho +\frac{a}{2} B_\mu Y^{\mu\nu} B_\nu \nn
&\simeq& -\frac{1}{2a} \chi_\mu Y^{\mu\nu} \chi_\nu
+ i \bar c_\mu Y^{\mu\nu} [ g_{\nu\rho} \Box +(2b+1)\nabla_\nu \nabla_\rho +\br_{\nu\rho}
] c^\rho,
\label{gfgh}
\ena
where the auxiliary field $B_{\mu}$ is integrated out in the last line.
Here
\bea
\chi_\mu &\equiv& \nabla^\la h_{\la\mu} + b \nabla_\mu h, \nn
Y_{\mu\nu} &\equiv& \bg_{\mu\nu} \Box+ c \nabla_\mu \nabla_\nu - d \nabla_\nu \nabla_\mu.
\ena
where $a$, $b$, $c$ and $d$ are gauge parameters.
We choose them such that the non-minimal four derivative terms
$\nabla_\mu \nabla_\nu \nabla_\a \nabla_\b$, $\bg_{\mu\nu} \Box\nabla_\a \nabla_\b$
and $\bg_{\nu\b} \nabla_\mu \Box \nabla_\a$ cancel. This leads to the choice~\cite{BS2}
\bea
a= \frac{1}{\b+4\c},~~~
b= \frac{4\a+\b}{4(\c-\a)}, ~~~
c-d = \frac{2(\c-\a)}{\b+4\c}-1.
\ena
In order to simplify the gauge-fixing term, we will further choose $d=1$.
With these choices, the ghost operator is
\bea
\Delta_{gh}{}^\mu{}_\nu =\delta^\mu{}_\nu\Box+(1+2b)\nabla^\mu\nabla_\nu+\br^\mu{}_\nu\ .
\ena
In addition, we also impose the traceless condition on $h_{\mu\nu}$:
\bea
h_\mu{}^\mu=0.
\label{addgaugec}
\ena
This does not introduce any ghost contribution, so we can simply impose this.

Then, the quadratic terms in the action can be written in the form
$h^{\mu\nu} {\cal K}_{\mu\nu,\a\b} h^{\a\b}$,
where
\bea
{\cal K}=K \Box^2 + D_{\rho\la} \nabla^\rho \nabla^\la + W.
\ena
The explicit forms of the coefficients are~\cite{OP}
\bea
(K)_{\mu\nu,\a\b} = \frac{\b+4\c}{4} \Big(\bg_{\mu\a}\bg_{\nu\b}
+\frac{4\a+\b}{4(\c-\a)}\bg_{\mu\nu}\bg_{\a\b} \Big),
\label{k}
\ena
\bea
(D_{\rho\la})_{\mu\nu,\a\b} \hs{-2}&=& \hs{-2}
-2\c \bg_{\nu\b}\br_{\a\rho\la\mu}
+4\c \bg_{\rho\nu} \br_{\la\a\mu\b}
+(\b + 3\c) \bg_{\rho\la} \br_{\mu\a\nu\b}
-(2\b+4\c) \bg_{\a\rho}\bg_{\nu\b} \br_{\mu\la}
\nn && \hs{-2}
 -2\c \bg_{\nu\b} \br_{\mu\a} \bg_{\rho\la}
+ \b \bg_{\mu\nu}\bg_{\b\rho} \br_{\a\la}
-2\a \bg_{\a\rho} \bg_{\b\la} \br_{\mu\nu}
+2\a \bg_{\mu\nu} \bg_{\rho\la} \br_{\a\b}
+2 \c \bg_{\mu\nu} \br_{\a\rho\la\b}
\nn && \hs{-2}
+ \frac{\a}{2}\br (\bg_{\mu\a} \bg_{\nu\b} \bg_{\rho\la}
- \bg_{\mu\nu} \bg_{\a\b} \bg_{\rho\la}
- 2 \bg_{\nu\b} \bg_{\mu\rho} \bg_{\a\la}
+ 2 \bg_{\mu\nu} \bg_{\a\rho} \bg_{\b\la})
\nn && \hs{-2}
+2\c \bg_{\nu\rho} \bg_{\b\la} \br_{\mu\a}
+\Big(\frac{\b}{2}+\c \Big) \bg_{\mu\a} \bg_{\nu\b} \br_{\rho\la}
-\frac{\b}{4} \bg_{\mu\nu} \bg_{\a\b} \br_{\rho\la},
\label{d}
\ena
\bea
(W)_{\mu\nu,\a\b} \hs{-2}&=& \hs{-2}
\frac32 \c \bg_{\nu\b} \br_\mu{}^{\rho\la\s} \br_{\a\rho\la\s} %- \br_{\a\s\rho\la})
+ 4\c \br_{\rho\a\mu\la} \br_{\nu\b}{}^{\rho\la}
- \c \br_{\rho\a\mu\la} \br^\rho{}_{\nu\b}{}^\la
+(\b+5\c) \br_{\rho\mu\la\nu} \br^\rho{}_\a{}^\la{}_\b
\nn && \hs{-2}
+ 6\c \br^\rho_\mu \br_{\rho\a\b\nu}
+ \Big(\frac{\b}{2}+\c \Big) \br_{\mu\a} \br_{\nu\b}
+ \a\br \Big(\frac12\br_{\mu\a\nu\b}
+ \frac32\bg_{\b\nu}\br_{\mu\a} -  \bg_{\mu\nu}\br_{\a\b}\Big)
\nn && \hs{-2}
+ \a \br_{\mu\nu} \br_{\a\b}
+ \frac{1}{8} (\a \br^2 +\b \br_{\rho\la}^2 +\c \br_{\rho\la\s\tau}^2 )
(\bg_{\mu\nu} \bg_{\a\b}-2\bg_{\mu\a} \bg_{\nu\b} )
+\Big(\frac{5}{2}\b+4\c \Big) \bg_{\nu\b} \br_{\mu\rho} \br^\rho_\a
\nn && \hs{-2}
- \c \bg_{\mu\nu} \br_\a{}^{\rho\la\s} \br_{\b\rho\la\s}
- \b \bg_{\a\b} \br_{\mu\rho} \br^\rho_\nu
- (\b+4\c) \bg_{\nu\b} \br^{\rho\la} \br_{\mu\rho\a\la},
\label{w}
\ena
where we have dropped terms with two derivatives acting on a background curvature,
and the symmetrization $\a\lra\b$ and $\mu\lra\nu$ and $(\mu,\nu)\leftrightarrow (\alpha,\beta)$
should be understood.

To impose the gauge condition~\p{addgaugec}, we should multiply the projection operator
\bea
P_{\mu\nu,\a\b}=\d_{\mu\nu,\a\b}-\frac1D \bg_{\mu\nu}\bg_{\a\b}.
\label{proj}
\ena
After this, the above kinetic operators are transformed into
\bea
(\tilde K)_{\mu\nu,\a\b} \hs{-2}&=& \hs{-2} (PKP)_{\mu\nu,\a\b} = \frac{\b+4\c}{4} P_{\mu\nu,\a\b},
\nn
(\tilde D_{\rho\la})_{\mu\nu,\a\b} \hs{-2}&=& \hs{-2} (PD_{\rho\la}P)_{\mu\nu,\a\b},
\nn
(\tilde W)_{\mu\nu,\a\b} \hs{-2}&=& \hs{-2} (PWP)_{\mu\nu,\a\b}.
\ena
As a final step, we factorize the tensor $\tilde K$ in the operator ${\cal K}$.
Since the projector $P$ in \p{proj} plays the role of the identity matrix, we find
\bea
{\cal K}= \tilde K{\cal H}\ ;\qquad
{\cal H}=\Box^2 + V_{\rho\la} \nabla^\rho \nabla^\la + U,
\label{hami}
\ena
where
\bea
V_{\rho\la} = \frac{4}{\b+4\c} \tilde D_{\rho\la}, ~~~
U = \frac{4}{\b+4\c} \tilde W.
\label{diffop}
\ena
The form of the coefficients $V_{\rho\lambda}$ and $U$ and their traces
are reported in appendix~\ref{sec:linear}.

\section{Quadratic expansion of the action for conformal theory
in the nonlinear exponential parametrization}

If we use the exponential type parametrization for the fluctuation~\p{nonlinear},
we find that the quadratic terms are slightly different.
We can easily derive the difference by noting that \p{nonlinear} gives
\bea
g_{\mu\nu} &=& \bg_{\mu\nu} + h_{\mu\nu} + \frac12 h_{\mu\la} h^\la{}_\nu + \ldots, \\
g^{\mu\nu} &=& \bg^{\mu\nu} - h^{\mu\nu} + \frac12 h^{\mu\la} h_\la{}^\nu + \ldots.
\ena
Thus if we restrict our attention only to the second order terms, they are
generated from the linear term in the expansion of the action, which is
\bea
{\cal L}^{(1)}
&=& \a \Big( \frac12 \br^2 h -2 \br \br_{\mu\nu} h^{\mu\nu} \Big)
+ \b \Big( \frac12 \br_{\mu\nu}^2 h - 2 \br^{\mu\nu} \br_{\a\mu\b\nu} h^{\a\b} \Big) \nn
&& +\  \c \Big( \frac12 \br_{\mu\nu\rho\la}^2 h
 - 2 \br_{\mu\nu\rho\la} \br_\a{}^{\nu\rho\la} h^{\mu\a} \Big),
\ena
up to terms which do not contribute to our results.
These terms generate second order terms by the replacement
\bea
h_{\mu\nu} \to \frac12 h_{\mu\la} h^\la{}_\nu.
\ena
This does not affect $K$ in \p{k} and $D_{\rho\la}$ in \p{d}, but modifies $W$ in \p{w} to
\bea
(\tilde W)_{\mu\nu,\a\b} \hs{-2}&=& \hs{-2}
\frac12 \c \bg_{\nu\b} \br_\mu{}^{\rho\la\s} \br_{\a\rho\la\s} %- \br_{\a\s\rho\la})
+ 4\c \br_{\rho\a\mu\la} \br_{\nu\b}{}^{\rho\la}
- \c \br_{\rho\a\mu\la} \br^\rho{}_{\nu\b}{}^\la
+(\b+5\c) \br_{\rho\mu\la\nu} \br^\rho{}_\a{}^\la{}_\b
\nn && \hs{-2}
+ 6\c \br^\rho_\mu \br_{\rho\a\b\nu}
+ \Big(\frac{\b}{2}+\c \Big) \br_{\mu\a} \br_{\nu\b}
+ \a\br \Big(\frac12\br_{\mu\a\nu\b}
+ \frac12 \bg_{\b\nu}\br_{\mu\a} -  \bg_{\mu\nu}\br_{\a\b}\Big)
\nn && \hs{-2}
+ \a \br_{\mu\nu} \br_{\a\b}
+ \frac{1}{8} (\a \br^2 +\b \br_{\rho\la}^2 +\c \br_{\rho\la\s\tau}^2 )
\bg_{\mu\nu} \bg_{\a\b}
+\Big(\frac{5}{2}\b+4\c \Big) \bg_{\nu\b} \br_{\mu\rho} \br^\rho_\a
\nn && \hs{-2}
- \c \bg_{\mu\nu} \br_\a{}^{\rho\la\s} \br_{\b\rho\la\s}
- \b \bg_{\a\b} \br_{\mu\rho} \br^\rho_\nu
- 2 ( \b + 2\c ) \bg_{\nu\b} \br^{\rho\la} \br_{\mu\rho\a\la},
\label{winexp}
\ena
{\em Note that the additional terms are proportional to the field equation because
the variation of the linear term gives precisely the field equation.}
After imposing the gauge condition~\p{addgaugec}, by multiplying the projector~\p{proj},
we get modified $U$ which is given in appendix~\ref{exp1}.

\section{Quadratic expansion of the action with Einstein and cosmological terms
in the nonlinear exponential parametrization}

Here let us consider the action
\bea
S &=& \int d^D x \sqrt{- g} \Big[ \frac{1}{\kappa^2} (R-2 \Lambda)
+ \a R^2 +\b R_{\mu\nu}^2 + \c R_{\mu\nu\rho\la}^2 \Big] .
\ena
We use the exponential parametrization~\p{nonlinear} in order to see if the result
changes from when we use the linear split~\p{linear} \cite{OP}.
In this case, our $K$ remains the same as \p{k}, but we do not have projection.
This gives the same $D_{\rho\la}$ as in our previous paper~\cite{OP} and modifies $W$ in \p{w} to
\bea
(D_{\rho\la})_{\mu\nu,\a\b} \hs{-2}&=& \hs{-2}
-2\c \bg_{\nu\b}\br_{\a\rho\la\mu}
+4\c \bg_{\rho\nu} \br_{\la\a\mu\b}
+(\b + 3\c) \bg_{\rho\la} \br_{\mu\a\nu\b}
-(2\b+4\c) \bg_{\a\rho}\bg_{\nu\b} \br_{\mu\la} \nn
&& \hs{-2} -2\c \bg_{\nu\b} \br_{\mu\a} \bg_{\rho\la}
+ \b \bg_{\mu\nu}\bg_{\b\rho} \br_{\a\la}
-2\a \bg_{\a\rho} \bg_{\b\la} \br_{\mu\nu}
+2\a \bg_{\mu\nu} \bg_{\rho\la} \br_{\a\b}
+2 \c \bg_{\mu\nu} \br_{\a\rho\la\b} \nn
&& \hs{-2}
+ \Big(\frac{\a}{2}\br +\frac{1}{4\kappa^2} \Big) (\bg_{\mu\a} \bg_{\nu\b} \bg_{\rho\la}
- \bg_{\mu\nu} \bg_{\a\b} \bg_{\rho\la}
- 2 \bg_{\nu\b} \bg_{\mu\rho} \bg_{\a\la}
+ 2 \bg_{\mu\nu} \bg_{\a\rho} \bg_{\b\la}) \nn
&& \hs{-2}  +2\c \bg_{\nu\rho} \bg_{\b\la} \br_{\mu\a}
+\Big(\frac{\b}{2}+\c \Big) \bg_{\mu\a} \bg_{\nu\b} \br_{\rho\la}
-\frac{\b}{4} \bg_{\mu\nu} \bg_{\a\b} \br_{\rho\la},
\ena
\bea
(W)_{\mu\nu,\a\b} \hs{-2}&=& \hs{-2}
\frac12 \c \bg_{\nu\b} \br_\mu{}^{\rho\la\s} \br_{\a\rho\la\s} %- \br_{\a\s\rho\la})
+ 4\c \br_{\rho\a\mu\la} \br_{\nu\b}{}^{\rho\la}
- \c \br_{\rho\a\mu\la} \br^\rho{}_{\nu\b}{}^\la
+(\b+5\c) \br_{\rho\mu\la\nu} \br^\rho{}_\a{}^\la{}_\b
\nn && \hs{-2}
+ 6\c \br^\rho_\mu \br_{\rho\a\b\nu}
+ \Big(\frac{\b}{2}+\c \Big) \br_{\mu\a} \br_{\nu\b}
+ \Big(\a\br+\frac{1}{2\kappa^2}\Big) \Big(\frac12\br_{\mu\a\nu\b}
+ \frac12 \bg_{\b\nu}\br_{\mu\a} -  \bg_{\mu\nu}\br_{\a\b}\Big)
\nn && \hs{-2}
+ \a \br_{\mu\nu} \br_{\a\b}
+ \frac{1}{8} \Big(\a \br^2 +\b \br_{\rho\la}^2 +\c \br_{\rho\la\s\tau}^2
+\frac{1}{\kappa^2} (\br -2 \Lambda) \Big) \bg_{\mu\nu} \bg_{\a\b}
- \b \bg_{\a\b} \br_{\mu\rho} \br^\rho_\nu
\nn && \hs{-2}
+\Big(\frac{5}{2}\b+4\c \Big) \bg_{\nu\b} \br_{\mu\rho} \br^\rho_\a
- \c \bg_{\mu\nu} \br_\a{}^{\rho\la\s} \br_{\b\rho\la\s}
- (2\b+4\c) \bg_{\nu\b} \br^{\rho\la} \br_{\mu\rho\a\la}.
\ena
We get modified $U$ which is given in appendix~\ref{exp2}.
We see that the result becomes simpler than the other parametrization~\cite{OP}.

We are now ready to discuss the beta functions in the renormalization group (RG) equation.

%%%%%%%%%%%%%%%%%%%%%%%%%%%%%%%%%%%%%%%%%%%%%%%%%%%%%%%%%%%%%%%%%%%%%%%%%%%%%%%%%%%%%
\section{Derivation of beta functions from functional renormalization group equation}
%%%%%%%%%%%%%%%%%%%%%%%%%%%%%%%%%%%%%%%%%%%%%%%%%%%%%%%%%%%%%%%%%%%%%%%%%%%%%%%%%%%%%

In the Wilsonian RG, we consider the effective action $\G_k$
describing physical phenomena at momentum scale $k$, which can be regarded as
the lower limit of the functional integration and the infrared cutoff.
The dependence of the effective action on $k$ gives the RG flow,
which can be written as a FRGE~\cite{wetterich} having on the r.h.s.
a trace of functions of the kinetic operators.

Up to this point, we have considered the action in Minkowski space.
In the following derivation of beta functions, we make Euclideanization.

In our quadratic action, we have the three operators: ${\cal H}$
acting on the graviton $h_{\mu\nu}$, the ghost operator $\Delta_{gh}$
and the third ghost operator $Y^{\mu\nu}$.
Let us choose cutoffs for the graviton, ghost and third ghost to be functions
of these operators, respectively:
$K R_k({\cal H})$ for the graviton,
$R_k(\Delta_{gh})$ for the ghosts and
$R_k(Y)$ for the third ghost.
The FRGE says that
\bea
\dot\Gamma_k=\frac{1}{2}\mbox{Tr}\frac{\dot R_k({\cal H})}{P_k({\cal H})}
-\mbox{Tr}\frac{\dot R_k(\Delta_{gh})}{P_k(\Delta_{gh})}
-\frac{1}{2}\mbox{Tr}\frac{\dot R_k(Y)}{P_k(Y)},
\ena
where we define $P_k(z)=z+R_k(z)$, and the dot represents the derivative with
respect to $\ln k$.
One can obtain the beta functions of $\alpha$, $\beta$, $\gamma$
by calculating the terms in the r.h.s. proportional to
$\int dx\sqrt{g}\br^2$,
$\int dx\sqrt{g}\br_{\mu\nu}\br^{\mu\nu}$ and
$\int dx\sqrt{g}\br_{\mu\nu\rho\sigma}\br^{\mu\nu\rho\sigma}$.

We can compute the r.h.s. using the following
general formulas for the trace of a function of an operator.
Calling $\tilde W$ the Laplace transform of $W$, we have for
a differential operator of order $p$ in $D$ dimensions:
\bea
\mathrm{Tr}[W(\Delta)] &=& \sum_n W(\lambda_n)=
\sum_n \int_{0}^{\infty}\!\! ds \, e^{-\lambda_n s} \tilde{W}(s)
=\int_{0}^{\infty}\!\! ds\,\tilde{W}(s)\,\mathrm{Tr}\, e^{-s\Delta}
\nonumber
\\
&=&
\sum_{n=0}^{\infty}B_{2n}(\Delta)
\int_{0}^{\infty}\!\! ds \, \tilde{W}(s) s^{-\frac{D}{p}+\frac{2n}{p}} =
\sum_{n=0}^{\infty} B_{2n}(\Delta)\, Q_{\frac{D-2n}{p}}(W),
\ena
where $B_{2n}$ are the coefficients appearing in the expansion of the heat kernel of the operator
\bea
\mathrm{Tr}\, e^{-s\Delta} =
\sum_{n=0}^{\infty}B_{2n}(\Delta)s^{-\frac{D}{p}+\frac{2n}{p}},
\ena
and the $Q$-functionals are given (for $m>0$) by
\bea
Q_m(W)=\int_{0}^{\infty}\!\! ds \, \tilde{W}(s) s^{-m}
=\frac{1}{\Gamma(m)}\int_{0}^{\infty}\!\! dz z^{m-1} W(z).
\ena
The last form is the more useful one.
(It is obtained more easily going from right to left.
Insert the Laplace expansion of $W$ in the r.h.s.,
exchange the order of the integrations over $s$ and $z$, and then use the integral
representation of the Gamma function.)
For $m=0$, one has $Q_0(W)=W(0)$.

With this formula the FRGE, expanded up to terms quadratic in curvature, is
\bea
\label{master}
\dot\Gamma_k
=
\frac{1}{2}B_4({\cal H})Q\left(4,(D-4)/4\right)
-B_4(\Delta_{gh})Q\left(2,(D-4)/2\right)
-\frac{1}{2}B_4(Y)Q\left(2,(D-4)/2\right).
\ena
We have to calculate the $Q$-functionals
$Q\left(p,m\right)=Q_m\left(\frac{\dot R_k}{P_k}\right)$, for an operator of order $p$.
For convenience, we choose the cutoff profile \cite{Litim}
$R_k(z)=(k^p-z)\theta(k^p-z)$, where $z$ is a differential operator of order $p$.
Define $z=yk^p$ and then we have
\bea
R_{k}(z)&=&(k^p-z)\theta (k^p-z) = k^p (1-y)\theta(1-y), \\
\dot{R}_{k}(z)&=&pk^p\theta (k^p-z) = pk^p\theta(1-y), \\
P_k (z)&=& z+R_{k}(z) = k^p\quad \mbox{for }\ z<k^p, \\
\frac{\dot R_k}{P_k}&=&p\,\theta(1-y).
\ena
For $m\geq 0$, we find
\bea
Q(p,m) &=&
\frac{1}{\Gamma(m)}\int_{0}^{\infty}\!\! dz z^{m-1} \frac{\dot{R}_{k}(z)}{P_k (z)}
= \frac{k^{mp}}{\Gamma(m)}\int_{0}^{\infty}\!\! dy y^{m-1} p\theta(1-y)
\nonumber \\
&=&  \frac{pk^{mp}}{\Gamma(m)}\int_0^1\! dy y^{m-1}
=  \frac{pk^{mp}}{\Gamma(m+1)}.
\label{charlie}
\ena

Next we list the necessary heat kernel coefficients.
{}From \cite{Gusynin1990}, we have
\bea
B_4({\cal H})
\hs{-2}&=& \hs{-2}
\frac{1}{(4\pi)^{D/2}}\frac{\Gamma(D/4)}{2\Gamma((D-2)/2)}
\int d^Dx\,\mathrm{tr}\Bigl[
\frac{\hat 1}{90} \br_{\rho\la\s\tau}^2 -\frac{\hat 1}{90} \br_{\rho\la}^2
+\frac{\hat 1}{36}\br^2 +\frac{1}{6} {\cal R}_{\rho\la}{\cal R}^{\rho\la} \nn
&& \hs{-7} - \frac{2}{D-2} U - \frac{1}{6(D-2)} (2 \br_{\rho\la}V^{\rho\la}
- \br V^\rho{}_\rho) + \frac{1}{4(D^2-4)} (V^\rho{}_\rho V^\la{}_\la
+ 2 V_{\rho\la} V^{\rho\la}) \Bigr],~~
\label{grav}
\ena
where $\hat 1$ is the identity defined in \p{id} and
${\cal R}_{\rho\la}$ is the commutator of the covariant derivatives acting
on the tensor $h^{\a\b}$: ${\cal R}_{\rho\la} = [\nabla_\rho, \nabla_\la ]$.
Collecting, we find
\bea
B_4({\cal H})
\hs{-2}&=&\hs{-2}
\frac{1}{(4\pi)^{D/2}}\frac{\Gamma(D/4)}{2\Gamma((D-2)/2)}
\int d^Dx\, \Biggl[
\br_{\mu\nu\rho\la}^2\left(\frac{(D-1)(D+2)}{180}-\frac{D+2}{6}
-\frac{2A_1}{D-2}+\frac{12D_1}{D^2-4}\right)
\nn
&&
-\br_{\mu\nu}^2\left(\frac{(D-1)(D+2)}{180}+\frac{2A_2}{D-2}
+\frac{C_1}{3(D-2)}-\frac{12D_2}{D^2-4}\right)
\nn
&&
+\br^2 \left(\frac{(D-1)(D+2)}{72}-\frac{2A_3}{D-2}+\frac{B_1}{6(D-2)}
-\frac{C_2}{3(D-2)}+\frac{12D_3}{D^2-4}\right)
\Biggr],
\label{fre}
\ena
where the constants $A_i, B_i,C_i$ and $D_i$ are defined in appendix~\ref{uv}.
Note that because we take the trace over the traceless symmetric tensor space,
we have to use tr$(\hat 1)=\frac{(D-1)(D+2)}{2}$, which is one of the differences
from the case in \cite{OP}.

{}From \cite{GK1999}, we have for the Euclidean operator
\bea
Y_{\mu\nu}=-\bg_{\mu\nu}\Box+\sigma_Y\nabla_\mu\nabla_\nu+\br_{\mu\nu},
\ena
with $\sigma_Y=1-2\frac{\gamma-\alpha}{\beta+4\gamma}$,
\bea
B_4(Y)
\hs{-2}&=&\hs{-2}
\frac{1}{(4\pi)^{D/2}}\int d^D x \sqrt{-\bg}
\Big[
\frac{D-16+(1-\sigma_Y)^{\frac{4-D}{2}}}{180} \br_{\mu\nu\rho\la}^2
\nn
&& \qquad\qquad
-\frac{D-91+(1-\sigma_Y)^\frac{4-D}{2}}{180} \br_{\mu\nu}^2
+\frac{D-13+(1-\sigma_Y)^\frac{4-D}{2}}{72} \br^2
\Big],
\ena
whereas for the Euclidean ghost operator
\bea
\Delta_{gh\mu\nu}=-\bg_{\mu\nu}\Box+\sigma_g\nabla_\mu\nabla_\nu-\br_{\mu\nu},
\ena
with $\sigma_g=-(1+2b)=-\left(1+2\frac{\beta+4\alpha}{4(\gamma-\alpha)}\right)$,
\bea
B_4(\Delta_{gh})
\hs{-2}&=&\hs{-2}
\frac{1}{(4\pi)^{D/2}}\int d^D x \sqrt{-\bg} \Big[
\frac{D-16+(1-\sigma_g)^{\frac{4-D}{2}}}{180} \br_{\mu\nu\rho\la}^2
\nn
&&
-\frac{(1-\sigma_g )^{-D/2}}{180D(D^2-4)\sigma_g}\Big(
D(D^2-4)\sigma_g^3
-2D(D+2)(D+58)\sigma_g^2
\nn
&&
+(D+2)(D^2+118D+720)\s_g
-1440D
\nn
&&
+(1-\sigma_g )^{D/2}
\{ (D^4-91D^3+596D^2-596D-1440 )\s_g+1440D \}
\Big)\br_{\mu\nu}^2
\nn
&&
+\frac{(1-\sigma_g )^{-D/2}}{72D(D^2-4)\sigma_g}\Big(
D(D^2-4)\sigma_g^3
-2D(D+2)(D+10)\sigma_g^2
\nn
&&
+(D+2)(D^2+22D+144) \s_g
-576
\nn
&&
+ (1-\sigma_g )^{D/2} \{(D^4+11 D^3-28D^2+52D-288)\s_g+576 \}
\Big)
\br^2
\Big],
\ena
The $B_4$ agree with the formulas in \cite{BS2} for $D=4$, but the dependence on $D$ is more
complicated than appears there.

Substituting the heat kernel coefficients in Eq.~(\ref{master}),
and extracting the coefficients of $\br^2$, $\br_{\mu\nu}\br^{\mu\nu}$,
and $\br_{\mu\nu\rho\sigma}\br^{\mu\nu\rho\sigma}$,
we obtain the beta functions of $\alpha$, $\beta$ and $\gamma$.
It turns out that the scalar curvature squared $\br^2$ is absent,
so our result is consistent with conformal invariance.

\section{Beta functions in four dimensions}

In this section, we summarize our results for beta functions in four dimensions.

\subsection{Conformal gravity in the linear split}

In four dimensions, the couplings $\alpha$ and $\beta$
(or equivalently $\lambda$ and $\rho$) are all dimensionless.
We define
\bea
\t \equiv \frac{\la}{\rho},
\ena
and consider beta functions for the couplings.
For $D=4$ and linear parametrization in Sec.~3, we have the traces in \ref{sec:linear}.
Using these results in the formulae in Sec.~4, we find
\bea
\b_\la &=& -\frac{1}{(4\pi)^2}\frac{199}{15} \la^2, \nn
\b_\rho &=& \frac{1}{(4\pi)^2}\frac{87}{20} \rho^2,
\ena
in agreement with \cite{BS1}. We also find
\bea
\b_\t = \frac{1}{(4\pi)^2}\frac{261-796 \t}{60} \la.
\ena
The fixed point of $\t$ is $\t_*=\frac{261}{796}=0.32789$.
The fixed point value that we found in our previous paper~\cite{OP} as a candidate
for four-dimensional conformal theory is $\t=0.325 296$.
As we will see, this value itself is closer to the fixed point in gravity theory
without conformal invariance. It seems that the main difference comes from the difference
in the contribution of the identity operator noted below Eq.~\p{fre}.
This in turn means the difference in the contribution of the trace part of $h_{\mu\nu}$.

\subsection{Conformal theory in the exponential parametrization}

On the other hand, it is interesting to check how the results change
if we use the nonlinear exponential parametrization in Sec.~4.
The trace of $U$ drastically simplifies as given in appendix~\ref{exp1}.
We find beta functions for these coefficients for $D=4$:
\bea
\b_\la &=& -\frac{1}{(4\pi)^2} \frac{199}{15} \la^2, \nn
\b_\rho &=& -\frac{1}{(4\pi)^2} \frac{87}{20} \rho^2, \nn %\frac{\la^2}{\t^2}, \nn
\b_\t &=& \frac{1}{(4\pi)^2} \frac{261-796\t}{60} \la,
\ena
To somewhat our surprise, we find that these results agree with the above beta functions
with the separate parametrization even though the traces of
$V$ and $U$ are so different.

\subsection{Quadratic theory with the Einstein and cosmological terms}

We have also examined the beta functions in the presence of Einstein and cosmological terms
in the exponential parametrization.
We use the couplings $\tilde\Lambda=\Lambda/k^2$, $\tilde G=G\,k^2$.
The results for the traces are summarized in
appendix~\ref{exp2}. In this case, we find that if we set the scalar curvature squared
$R^2$ to zero from the outset in the action~\p{action}, we get singular beta functions.
So we first set the coefficients in the action~\p{action1} as
\bea
\a = -\frac{1}{\rho}+\frac{1}{\xi}+\frac{1}{6\la}, ~~
\b = \frac{4}{\rho} -\frac{1}{\la}, ~~
\c = -\frac{1}{\rho} +\frac{1}{2\la}.
\label{coe1}
\ena
We also define
\bea
\xi = -\frac{(D-1)\la}{\omega}.
\ena
We obtain results that should be compared with those in our previous paper~\cite{OP}.

The beta functions are found to be
\bea
\b_\la &=& -\frac{1}{(4\pi)^2} \frac{133}{10} \la^2, \nn
\b_\rho &=& -\frac{1}{(4\pi)^2} \frac{196}{45} \rho^2, \nn
\b_\xi &=& -\frac{1}{(4\pi)^2} \frac{5(1+12 \omega+8 \omega^2)}{4 \omega^2} \la^2, \nn
\b_\omega &=& -\frac{1}{(4\pi)^2} \frac{25+1098 \omega+200 \omega^2}{60} \la, \nn
\b_\t &=& \frac{1}{(4\pi)^2} \frac{7(56-171 \t)}{90} \la,
\ena
We find that these agree with the results in Ref.~\cite{OP}.
The fixed points of $\la$ and $\t$ are $\la_*=0$ and $\t_*=\frac{56}{171}=0.3275$,
respectively.

We also have the beta functions for the Newton and cosmological constants.
For simplicity we omit tildes on these.
\bea
\b_\Lambda &=& -2\Lambda +\frac{3+34 \omega+ 40 \omega^2}{12(4\pi)^2\omega}\la\Lambda
-  \frac{171+298\omega +152\omega^2+16\omega^3}{36\pi (1+\omega)} G\Lambda \nn
&& +\frac{283+664\omega+204\omega^2-128\omega^3-32\omega^4}{144 \pi (1+\omega)^2} G
- \frac{1+10\omega}{4 (4\pi)^2\omega}\la +\frac{1+20\omega^2}{64 (4\pi)^3 \omega^2 G}\la^2 , \nn
\b_G &=& 2 G - \frac{5-26\omega-40\omega^2}{12(4\pi)^2\omega} \la G
 - \frac{171+298\omega+152\omega^2+16\omega^3}{36 \pi (1+\omega)} G^2.
\ena
The beta functions for $G$ and $\Lambda$ in the linear split~\p{linear}
were found to be~\cite{OP}
\bea
\beta_{ \Lambda} & = &
-2\Lambda
+\frac{1+86\omega+40\omega^2}{12(4\pi)^{2}\omega}\lambda\Lambda
-\frac{171+298\omega+152\omega^2+16\omega^3}{36\pi (1+\omega)} G \Lambda \nn
&& +\frac{283+664\omega+204\omega^2-128\omega^3-32\omega^4}{144\pi(1+\omega)^2} G
-\frac{1+10\omega}{4(4\pi)^{2}\omega}\lambda
+\frac{1+20\omega^2}{64(4\pi)^3\omega^2  G}\lambda^2 \ ,
\nn
\beta_{ G} & = & 2  G-\frac{3+26\omega-40\omega^2}{12(4\pi)^{2}\omega}\lambda G
-\frac{171+298\omega+152\omega^2+16\omega^3}{36\pi (1+\omega)} G^2
\ ,
\label{fleq1}
\ena

Thus we find that the beta functions for dimensionful couplings are different
though those for the dimensionless couplings remain the same. This suggests the universality
of the latter beta functions. We can also see that only the second terms of both the beta
functions for $G$ and $\Lambda$ are different,
but these terms vanish at the fixed point of $\lambda_*=0$, so the fixed points are
not affected by the difference. This suggests that our results are parametrization independent.
For a picture of the flow in the $\Lambda$-$G$ plane, see \cite{OP}.

\section{Conclusions}

In this paper, we have studied the beta functions for conformal gravity in the linear split
and exponential parametrization of the metric fluctuation and found that they agree with
each other. This is an indication that these results are universal.
This study was partly motivated by our study of fixed points in Einstein theory with
cosmological and quadratic curvature terms~\cite{OP} where we find an extra fixed point
in addition to those already known. We conjectured that this might correspond to
conformal gravity in four dimensions because the coefficient $\frac{1}{\xi}$
for the scalar curvature squared vanishes at the fixed point. However, our analysis
indicates that this does not seem to be the case since the fixed point value are different
for the conformal gravity independently of the parametrization.
The contribution from the trace modes of the metric seems to make the crucial difference.

To check the parametrization dependence of this approach, we have also examined
the beta functions for the Einstein theory with cosmological constant and quadratic terms.
We have found that the beta functions for dimensionless couplings are the same,
but those for dimensionful couplings are slightly different. This suggests again
that the beta functions for dimensionless couplings are universal, but those for
dimensionful couplings are not. However, we find that the fixed points are same
because the difference disappears at the fixed point.
These are physically reasonable results.

As mentioned in the introduction, there is an issue of gauge-dependence in these results.
It would be very interesting to study the gauge and parametrization independence and other
issues, to which we hope to return in a separate publication~\cite{OP2}.

\section*{Acknowledgment}

This work was supported in part by the Grant-in-Aid for
Scientific Research Fund of the JSPS (C) No. 24540290.

\appendix
\section{$U$, $V$ and their traces}
\label{uv}

\subsection{Linear split}
\label{sec:linear}

The tensors $U$ and $V$ are given in (\ref{diffop}) for the case of linear split~\p{linear}
after the traceless projector~\p{proj} is multiplied.
The explicit form of $U$ is given by
\bea
U \hs{-2}&=& \hs{-2} \frac{4}{\b+4\c} \Big[
\frac{1}{4}\Big( \a \br^2 + \b \br_{\rho\la}^2+\c \br_{\rho\la\s\tau}^2 \Big)
\Big(\frac{D+12}{D^2}\bg_{\mu\nu}\bg_{\a\b}-\bg_{\mu\a}\bg_{\nu\b}\Big)
+\frac{3}{2D}\a(D \bg_{\b\nu}\br_{\mu\a}
\nn && \hs{-2}
-2\bg_{\a\b}\br_{\mu\nu}
-2\bg_{\mu\nu}\br_{\a\b}) \br
+\frac{\a}2\br_{\mu\a\nu\b} \br
-2\frac{3\b+2\c}{D} \bg_{\a\b}\br_{\mu\rho}\br_\nu^\rho
+ \Big(\frac{\b}{2}+\c \Big) \br_{\mu\a} \br_{\nu\b}
\nn && \hs{-2}
+ \a \br_{\mu\nu} \br_{\a\b}
+\Big(\frac{5}{2}\b+4\c \Big) \bg_{\nu\b} \br_{\mu\rho} \br^\rho_\a
+\frac{4}{D}\c \br^{\rho\la} \br_{\rho\mu\la\nu} \bg_{\a\b}
- (\b+4\c) \bg_{\nu\b} \br^{\rho\la} \br_{\mu\rho\a\la}
\nn && \hs{-2}
+ 6\c \br^\rho_\mu \br_{\rho\a\b\nu}
+ 4\c \br_{\rho\a\mu\la} \br_{\nu\b}{}^{\rho\la}
-\frac{3}{D} \c \bg_{\a\b} \br_{\mu\rho\la\s} \br_\nu{}^{\rho\la\s}
- \c \br_{\rho\a\mu\la} \br^\rho{}_{\nu\b}{}^\la
\nn && \hs{-2}
+(\b+5\c) \br_{\rho\mu\la\nu} \br^\rho{}_\a{}^\la{}_\b
+ \frac32 \c \bg_{\nu\b} \br_\mu{}^{\rho\la\s} \br_{\a\rho\la\s}
- \frac{3}{D} \c \bg_{\mu\nu} \br_\a{}^{\rho\la\s} \br_{\b\rho\la\s} \Big].
\label{curm}
\ena
The expression for $V^{\rho\la}=(V^{\rho\la})_{\mu\nu,\a\b}$ is
\bea
V^{\rho\la} = \frac{4}{\b+4\c} \sum_{i=1}^{20} b_i \bk_i,
\label{expv}
\ena
where
\bea
&& \bk_1= \bg^{\rho\la} \bg_{\b(\nu} \br_{\mu)\a},~~~
\bk_2= \d_{\mu\nu,\a\b} \bg^{\rho\la},~~~
\bk_3= \bg^{\rho\la} \br_{\mu\a\nu\b},~~~
\bk_4= \d_{\nu\b}{}^{\rho\la} \br_{\mu\a}, \nn
&& \bk_5= \d_{\nu\b}{}^{\rho\la} \bg_{\mu\a},~~~~
\bk_6= \d_{\mu\nu,\a\b}\br^{\rho\la},~~~
\bk_7= \frac12 (\d_\nu^{(\rho} \br^{\la)}{}_{\a\b\mu} + \d_\b^{(\rho} \br^{\la)}{}_{\mu\nu\a}),\nn
&& \bk_8= \bg_{\nu\b}\d^{(\rho}_{(\mu} \br^{\la)}_{\a)}, ~~~
\bk_9 = \bg_{\nu\b} \br_{(\a}{}^{\rho\la}{}_{\mu)},~~~
\bk_{10}= \frac12 (\d_{\a\b}{}^{\rho\la}\br_{\mu\nu} + \d_{\mu\nu}{}^{\rho\la} \br_{\a\b}),\nn
&& \bk_{11} = \bg_{\mu\nu} \br_{\a}{}^{\rho\la}{}_{\b},~~~
\bk_{12} = \bg_{\a\b} \br_{\mu}{}^{\rho\la}{}_{\nu},~~~
\bk_{13} = \bg_{\mu\nu} \bg^{\rho\la} \br_{\a\b},~~~
\bk_{14} = \bg_{\a\b} \bg^{\rho\la} \br_{\mu\nu}, \nn
&& \bk_{15} = \bg_{\mu\nu} \d_{(\a}^{(\la} \br_{\b)}^{\rho)},~~~~
\bk_{16} = \bg_{\a\b} \d_{(\mu}^{(\la} \br_{\nu)}^{\rho)},~~~
\bk_{17}= \bg_{\mu\nu} \d_{\a\b}{}^{\rho\la},~~~
\bk_{18}= \bg_{\a\b} \d_{\mu\nu}{}^{\rho\la}, \nn
&& \bk_{19} = \bg_{\mu\nu} \bg_{\a\b} \bg^{\rho\la}, ~~~
\bk_{20} = \bg_{\mu\nu} \bg_{\a\b} \br^{\rho\la},
\ena
and
\bea
&& b_1= -2\c, ~~
b_2=\frac{\a}{2} \br, ~~
b_3 = \b+3\c, ~~
b_4 = 2\c, ~~
b_5 = -\a \br,~~
b_6 = \frac{\b}{2}+\c,~~~\nn
&&
b_7 = -4\c, ~~~
b_8 = -2\b-4\c, ~~~
b_9 = -2\c, ~~~
b_{10} = -2\a,~~~
b_{11} = \frac{4}{D}\c,~~~
b_{12} = \frac{4}{D}\c, \nn
&&
b_{13} = \frac{\a-\b-\c}{D}, ~~
b_{14} = \frac{\a-\b-\c}{D}, ~~
b_{15} = \frac{2\b}{D},~~
b_{16} = \frac{2\b}{D},~~
b_{17} = b_{18}=\frac{2\a}{D} \br,~~~  \nn
&&
b_{19}= \frac{-(D+6)\a+2(\b+\c)}{2D^2} \br, ~~~
b_{20}= -\frac{(D+4)}{2D^2}\b-\frac{D-4}{D^2} \c.
\label{v}
\ena
In deriving this result, one has to pay special attention to the symmetry
$(\mu,\nu)\leftrightarrow (\a,\b)$.

The following results are obtained using software Math Tensor run on the Mathematica.
It is important to realize that the indices are symmetrized. For example,
the indices $\rho$ and $\la$ on $V$ must be symmetrized in making products.

The trace of $U$ is given as
\bea
\mbox{tr }U = \d^{\mu\nu,\a\b} U_{\mu\nu,\a\b}
= A_1 \br_{\mu\nu\rho\la}^2 + A_2 \br_{\mu\nu}^2
+ A_3 \br^2,
\ena
where
\bea
A_1 &=& \frac{-24\c+2(3\b+13\c)D+5\c D^2-\c D^3}{2D(\b+4\c)}, \nn
A_2 &=& \frac{-24\b+2(4\a+5\b+12\c)D+5\b D^2-\b D^3}{2D(\b+4\c)}, \nn
A_3 &=& \frac{-24\a+2(3\a+\b+2\c)D+5\a D^2-\a D^3}{2D(\b+4\c)}.
\label{tracec1}
\ena
For $D=4$, these reduce to
\bea
A_1 = 3,~~~
A_2 = \frac{4(\a+\b+3\c)}{\b+4\c},~~~
A_3 = \frac{2\a+\b+2\c}{\b+4\c}.
\ena
Next,
\bea
\mbox{tr } (V^\rho_\rho \br)
= B_1 \br^2,
\ena
with
\bea
B_1 &=& \frac{(D+2) \{D(D-1)(D-2)\a+(D^2-7D+4)\b-2(D^2+2D+4)\c\} }{D(\b+4\c)}.
\ena
For $D=4$,
\bea
B_1 = \frac{12(3\a-\b-7\c)}{\b+4\c}.
\ena
Next,
\bea
\mbox{tr } (V^{\rho\la} \br_{\rho\la})
= C_1 \br_{\mu\nu}^2 + C_2 \br^2,
\ena
with
\bea
C_1 &=& \frac{8(\b-2\c)-2(4\a+3\b+6\c)D-(3\b+2\c)D^2+(\b+2\c)D^3}{D(\b+4\c)}, \nn
C_2 &=& \frac{4(3\a-\b-\c)-2(2\a+\b+3\c)D-(\a+4\c)D^2+\a D^3}{D(\b+4\c)}.
\ena
For $D=4$,
\bea
C_1 = \frac{8(-\a+\c)}{\b+4\c}, ~~~
C_2 = \frac{11\a-3\b-23\c}{\b+4\c}.
\ena
Finally
\bea
\frac{1}{48} \mbox{tr } (V^\rho_\rho V^\la_\la)
+\frac{1}{24} \mbox{tr } (V_{\rho_\la} V^{\rho\la})
= D_1 \br_{\mu\nu\rho\la}^2 + D_2 \br_{\mu\nu}^2 + D_3 \br^2,
\ena
where
\bea
D_1 &=& \frac{D^2(D+2)\b^2+2D(D+2)(3D+4)\b\c +(9D^3+48D^2+64D-64)\c^2}{4D(\b+4\c)^2},\nn
D_2 &=& \frac{1}{12D^2(\b+4\c)^2} \Big[8D^3(D+1)\a^2+32D(-D^2+D+4)\a\c +16(D^2+D-4)D\a\b \nn
&& +4(D+2)(7D^3+5D^2-16)\b\c +(3D^4+5D^3+30D^2+16D+32)\b^2 \nn
&& +4(D^5+14D^4+30D^3+2D^2-80D+32)\c^2 \Big],\nn
D_3 &=& \frac{1}{24D^2(\b+4\c)^2} \Big[D(D^5+D^4-2D^3-24D^2-24D+16)\a^2 \nn
&& +2(D^5-D^4-2D^3-20D^2-24D+96)\a\b \nn
&& -4(D^5+2D^4-8D^3-32D^2+32D+96)\a\c
+\{D(D-8)(D^2+D+6)-32\} \b^2 \nn
&& -4(D+2)(D^3+8D^2-6D+8) \b\c
 -4(D^4+17D^3+28D^2+4D-64)\c^2 \Big].
\ena
For $D=4$,
\bea
&& D_1 = 6,~~
D_2 = \frac{2(20\a^2+8\a\b+13\b^2-8\a\c+96\b\c+196\c^2)}{3(\b+4\c)^2},\nn
&& D_3 = \frac{43\a^2+10\a\b-7\b^2-46\a\c-66\b\c-109\c^2}{6(\b+4\c)^2}.
\ena
These results for quadratic curvature terms agree with those in \cite{BS1}.

\subsection{Nonlinear exponential parametrization}
\label{exp1}

When we use the exponential parametrization~\p{nonlinear}, the tensor $V$ is the same as
the linear split, but the tensor $W$ changes to \p{winexp}. The tensor $U$,
after the projection by \p{proj}, is then given by
\bea
U \hs{-2}&=& \hs{-2} \frac{4}{\b+4\c} \Big[
\frac{2}{D^2}\Big( \a \br^2 + \b \br_{\rho\la}^2+\c \br_{\rho\la\s\tau}^2 \Big)
 \bg_{\mu\nu}\bg_{\a\b} + \frac{\a}{2D}(D \bg_{\b\nu}\br_{\mu\a}
-4\bg_{\a\b}\br_{\mu\nu}
\nn && \hs{-2}
-4\bg_{\mu\nu}\br_{\a\b}) \br
+\frac{\a}2\br_{\mu\a\nu\b} \br
-2\frac{3\b+2\c}{D} \bg_{\a\b}\br_{\mu\rho}\br_\nu^\rho
+ \Big(\frac{\b}{2}+\c \Big) \br_{\mu\a} \br_{\nu\b}
+ \a \br_{\mu\nu} \br_{\a\b}
\nn && \hs{-2}
+\Big(\frac{5}{2}\b+4\c \Big) \bg_{\nu\b} \br_{\mu\rho} \br^\rho_\a
+\frac{2(\b+2\c)}{D} \br^{\rho\la} \br_{\rho\mu\la\nu} \bg_{\a\b}
- 2( \b+2\c) \bg_{\nu\b} \br^{\rho\la} \br_{\mu\rho\a\la}
\nn && \hs{-2}
+ 6\c \br^\rho_\mu \br_{\rho\a\b\nu}
+ 4\c \br_{\rho\a\mu\la} \br_{\nu\b}{}^{\rho\la}
-\frac{2}{D} \c \bg_{\a\b} \br_{\mu\rho\la\s} \br_\nu{}^{\rho\la\s}
- \c \br_{\rho\a\mu\la} \br^\rho{}_{\nu\b}{}^\la
\nn && \hs{-2}
+(\b+5\c) \br_{\rho\mu\la\nu} \br^\rho{}_\a{}^\la{}_\b
+ \frac{\c}{2} \bg_{\nu\b} \br_\mu{}^{\rho\la\s} \br_{\a\rho\la\s}
- \frac{2}{D} \c \bg_{\mu\nu} \br_\a{}^{\rho\la\s} \br_{\b\rho\la\s} \Big].
\ena
The trace is given as
\bea
\mbox{tr }U = \d^{\mu\nu,\a\b} U_{\mu\nu,\a\b}
= A_1 \br_{\mu\nu\rho\la}^2 + A_2 \br_{\mu\nu}^2
+ A_3 \br^2,
\ena
where
\bea
A_1 &=& \frac{-8\c+(3\b+10\c)D+\c D^2}{D(\b+4\c)}, \nn
A_2 &=& \frac{-8\b+2(2\a+\b+6\c)D+\b D^2}{D(\b+4\c)}, \nn
A_3 &=& \frac{-8\a+(\b+2\c)D+\a D^2}{D(\b+4\c)}.
\ena
Note that these are quite different from the corresponding ones~\p{tracec1} in the linear
split where they contain cubic terms in dimension $D$.
Nevertheless, it turns out that they agree with those in the linear split in four dimensions:
\bea
A_1 = 3,~~~
A_2 = \frac{4(\a+\b+3\c)}{\b+4\c},~~~
A_3 = \frac{2\a+\b+2\c}{\b+4\c}.
\ena

The rest of the traces of $V$ are the same as in the previous subsection~\ref{sec:linear}.

\subsection{Nonlinear exponential parametrization in Einstein-quadratic theory}
\label{exp2}

Here we consider the theory with quadratic curvature terms and the Einstein and
cosmological terms, as considered in our previous paper~\cite{OP}.
We use the exponential parametrization~\p{nonlinear}.
In this case, we do not impose the traceless condition, and we simply use
\bea
(K^{-1})_{\mu\nu}{}^{\a\b}
=\frac{4}{\b+4\c} (\d_{\mu\nu}{}^{\a\b} -\Omega \bg_{\mu\nu} \bg^{\a\b}),
\ena
with
\bea
\Omega = \frac{4\a+\b}\Sigma, ~~~
\Sigma \equiv 4(\c-\a)+D(4\a+\b).
\label{sigma}
\ena
After some work we find
\bea
(U)_{\mu\nu,\a\b}\!\!&=&\!\! \frac{4}{\b+4\c} \Bigg[
\frac12 \c \bg_{\nu\b}\br_\mu{}^{\rho\la\s} \br_{\a\rho\la\s}
-\c \br^\la{}_{\a\mu}{}^\rho \br_{\la\nu\b\rho}
+4\c \br_{\rho\a\mu\la} \br_{\nu\b}{}^{\rho\la} \nn
&& -3\c(\br_\mu^\s \br_{\s\a\nu\b} + \br^\s_\a \br_{\s\mu\b\nu})
+\Big(\frac{\b}{2}+\c \Big) \br_{\mu\a}\br_{\nu\b}
-\frac{\c}{2}\bg_{\a\b} \br_{\mu\rho\la\s} \br_\nu{}^{\rho\la\s} \nn
&& +\frac{\Omega_3}2  S^2 \bg_{\mu\nu}\bg_{\a\b} +\Big(\frac{\a}{2}\br +\frac{1}{4\kappa^2} \Big)
 (\br_{\mu\a\nu\b}+ \bg_{\nu\b} \br_{\mu\a} -\bg_{\a\b} \br_{\mu\nu}) \nn
&& +\Big(\frac{5}{2}\b+4\c \Big) \bg_{\nu\b} \br_{\mu\s} \br^{\s}_\a
+(\b+5\c) \br_{\rho\mu\la\nu} \br^\rho{}_\a{}^\la{}_\b
- \frac{\b}{2} \bg_{\a\b} \br_{\mu\s}\br^\s_\nu \nn
&& - 2\Omega_3 (\c \bg_{\mu\nu} \br_{\a\rho\la\s} \br_\b{}^{\rho\la\s}
+ \b \bg_{\mu\nu} \br_{\a\s}\br^\s_\b)
+ \a \br_{\mu\nu} \br_{\a\b}
-\Big( \frac{\Omega_1-2\Omega}{2\kappa^2} \nn
&& +2 \a \Omega_3 \br \Big) \bg_{\mu\nu} \br_{\a\b}
-\frac{1}{4\kappa^2}(\br-4\Lambda)\Omega \bg_{\mu\nu} \bg_{\a\b}
-2(\b+2\c) \bg_{\nu\b} \br^{\rho\la}\br_{\mu\rho\a\la} \Bigg],~~~~~~
\ena
where we have defined
\bea
S^2 = \a \br^2+\b \br_{\mu\nu}^2 + \c \br_{\mu\nu\rho\la}^2
+ \frac{1}{\kappa^2}( \br-2 \Lambda),
\ena
and
\bea
\Omega_1 = \frac{10\a+3\b+2\c}{\Sigma}, ~~~
%\Omega_2 = \frac{28\a\b-4\a\c+9\b^2+7\b\c}{4\Sigma}, ~~~
\Omega_3 = \frac{3\a+\b+\c}{\Sigma}, ~~~
\ena
with $\Sigma$ given in \p{sigma}.

The trace of $U$ is given as
\bea
\mbox{tr }U = \d^{\mu\nu,\a\b} U_{\mu\nu,\a\b}
= A_1 \br_{\mu\nu\rho\la}^2 + A_2 \br_{\mu\nu}^2
+ A_3 \br^2+A_4 \frac{\br}{\kappa^2}+A_5 \frac{\Lambda}{\kappa^2},
\ena
where
\bea
A_1 \hs{-2}&=& \hs{-2}
\frac{1}{(\b+4\c) \Sigma} \Big[12(D-1)\a\b+2(2D^2+17D-28)\a\c+3D \b^2 +(D^2+10D+4)\b\c \nn
&& \hs{20} +6(D+4)\c^2 \Big] , \nn
A_2 \hs{-2}&=& \hs{-2}
\frac{1}{(\b+4\c) \Sigma} \Big[16(D-1)\a^2+2(2D^2+3D-12)\a\b+16(3D-2)\a\c \nn
&& \hs{20} +(D^2+2D-8)\b^2 +2(9D-4)\b\c+48\c^2 \Big], \nn
A_3 \hs{-2}&=& \hs{-2}
\frac{1}{(\b+4\c) \Sigma} \Big[2(2D^2-3D-8)\a^2+(D^2+4D-12)\a\b+2(7D-12)\a\c \nn
&& \hs{20} +(\b+2\c)(D\b+4\c) \Big], \nn
A_4 \hs{-2}&=& \hs{-2}
\frac{1}{(\b+4\c) \Sigma} \frac{D-2}{2}\big[ 4D\a+(D+2)\b+8\c \big], \nn
A_5 \hs{-2}&=& \hs{-2} \frac{1}{(\b+4\c) \Sigma} 4D(\a-\c).
\ena

For the traces of $V$, we should use those in our previous paper~\cite{OP}.

\end{document}